\documentclass[referee]{aa}
\usepackage{graphics}
\usepackage{epsfig}

\def\Mplanet{M_{\rm planet}}

\def\msol{M_\odot}

\def\simgr{\,\hbox{\hbox{$ > $}\kern -0.8em \lower 1.0ex\hbox{$\sim$}}\,}
\def\simle{\,\hbox{\hbox{$ < $}\kern -0.8em \lower 1.0ex\hbox{$\sim$}}\,}
\def\beq{\begin{equation}}
\def\eeq{\end{equation}}
\def\dpartial#1#2{{{\partial {#1}} \over {\partial {#2}}}}

\def\as{a_{\rm S}}
\def\ms{M_{\rm S}}

\def\msol{M_\odot}

\def\simgr{\,\hbox{\hbox{$ > $}\kern -0.8em \lower 1.0ex\hbox{$\sim$}}\,}
\def\simle{\,\hbox{\hbox{$ < $}\kern -0.8em \lower 1.0ex\hbox{$\sim$}}\,}
\def\beq{\begin{equation}}
\def\eeq{\end{equation}}

\def\apjl{ApJ}                
\def\apjs{ApJS}               
\def\ao{Appl.Optics}          
\def\aap{A\&A}                

\def\({\left(}
\def\){\right)}
\def\<{\left<}
\def\>{\right>}

\begin{document}

\title{Modeling the Jovian subnebula: I - Thermodynamical conditions and migration of proto-satellites}

\author{Yann Alibert$^1$, Olivier Mousis$^{1,2}$ and Willy Benz$^1$}

   \offprints{Yann Alibert, \\
   e-mail: yann.alibert@phim.unibe.ch}

\institute{${}^1$  Physikalisches Institut, University of Bern, Sidlerstrasse 5, CH-3012 Bern, Switzerland \\
${}^2$ Observatoire de Besan\c con, CNRS-UMR 6091, BP 1615, 25010 BESANCON Cedex, France
}

\date{Received /Accepted}

\titlerunning{Modeling the Jovian subnebula I}

   \abstract{We have developed an evolutionary turbulent model of the Jovian
subnebula consistent with the extended core accretion formation models of Jupiter 
described by Alibert
et al. (2005b) and derived from Alibert et al. (2004,2005a). This model takes
into account the vertical structure of the subnebula, as well as the evolution
of the surface density as given by an $\alpha$-disk model  and is used to calculate the thermodynamical 
conditions in the subdisk, for different values of the viscosity parameter. 
We show that the Jovian subnebula evolves in two different phases during its 
lifetime. In the first phase, the subnebula is fed through its outer edge by 
the solar nebula as long as it has not been dissipated. In the second phase, 
the solar nebula has disappeared and the Jovian subdisk expands and gradually 
clears with time as Jupiter accretes the remaining material. We also 
demonstrate that early generations of satellites formed during the beginning 
of the first phase of the subnebula cannot survive in this environment and 
fall onto the proto-Jupiter. As a result, these bodies may contribute to 
the enrichment of Jupiter in heavy elements. Moreover, migration 
calculations in the Jovian subnebula allow us to follow the evolution of the 
ices/rocks ratios in the proto-satellites as a function of their migration 
pathways. By a tempting to reproduce the distance distribution of the Galilean 
satellites, as well as their ices/rocks ratios, we obtain some constraints 
on the viscosity parameter of the Jovian subnebula.

   \keywords{Planets and satellites: formation -- Solar system: formation}
   }

\authorrunning{Alibert Y. et al.}
\maketitle

\section{Introduction}

In  recent years, the conditions leading to the formation of regular satellite systems in the 
Saturnian and Jovian subnebulae have been studied in details by Mousis et al. 
(2002a), Canup \& Ward (2002), Mosqueira \& Estrada (2003a,b), and Mousis \& 
Gautier (2004). However, while these authors described the evolution of the 
gas-phase chemistry and/or the conditions of accretion of the regular satellites, 
none of them coupled the evolution of the subnebula to the last phases of the 
formation of Jupiter and Saturn.

For instance, Mousis et al. (2002a) and Mousis \& Gautier (2004) noted that the
accretion rates used in their subdisk models were substantially lower than those
calculated by Coradini et al. (1995) for the last stages of Saturn and 
Jupiter formation. According to them, adopting initial accretion rates in 
agreement with the Coradini et al. (1995) rates, would have been incompatible
with the assumption of geometrically thin disk on which their models were based. 
Moreover, they assumed subdisks models whose material (gas and gas-coupled solids) 
were not supplied by the solar nebula, in conflict with the actual scenarios of 
subnebula formation (Lubow et al. 1999, Magni \& Coradini 2004). Instead, they assumed 
that their subdisks were closed systems with a mass that can only decrease with
time, neglecting the fact that the giant planets continued to accrete gas from 
the solar nebula during a substantial fraction of time. 

Canup \& Ward (2002, hereafter referred to as CW02),
proposed an accretion disk model of the Jovian subnebula supported
by the simulations of solar nebula gas accretion through a gap onto a
Jupiter-like planet as described by Lubow et al. (1999). These authors
considered satellite formation in a low surface density circumjovian accretion disk,
resulting from the reduced gas inflow characterizing the last stages of planet's growth.
The regular satellite formation model of CW02 is therefore compatible
with the presence of water ice in the two Galilean icy satellites
as well as the apparent incomplete differentiation of Callisto.
Moreover, the reduced surface density naturally increases type
I migration timescales, thus making survival of satellites more likely.

Their model, however, depends on three unknown parameters, namely the
opacity inside the subdisk, the accretion rate of material (from the nebula
to the subnebula), and the viscosity parameter inside the subdisk.  Moreover,
the accretion rate is assumed to be time independent, implying a constant
supply of material from the solar nebula. Thus, the model of CW02 does not
take into account the progressive thining of the solar nebula and its
consequences on the structure and the time evolution of the Jovian
subnebula.

Finally, in their model of subnebula, although the temperature at the present
location of the two Galilean icy satellites is low enough to allow the presence 
of water ice in satellitesimals, it is too high to maintain the trapping of 
volatiles such as CO$_2$ and NH$_3$ under the form of hydrates or pure condensates 
(see Mousis et al. 2005 - hereafter refered paper II - for details). This is 
in conflict with evidences of the presence of these species in 
the two Galilean icy satellites (see Hibbitts et al. 2000, 2003; Mousis et al. 
2002b; Spohn \& Schubert 2003).

Finally, Mosqueira \& Estrada (2003a,b) considered stationary 
models of the Jovian and Saturnian subnebulae that were divided in two different 
zones. Each subnebula was assumed to be composed of an optically thick and 
potentially turbulent inner region located inside the planet's centrifugal radius, 
surrounded by an optically thin laminar extended disk. Since their models were
not explicitely tied to any giant planet formation calculations, it is difficult 
to evaluate whether such structures are natural outcomes of giant planet and their
subnebulae formation.

In a recent paper, Alibert et al. (2004) have proposed some 
new models of giant planet formation. These models include 
migration in solar nebula models whose evolution is ruled by 
viscous dissipation and photoevaporation.  Using these models, 
the authors showed that it is possible to form planets in a 
timescale well within observed disks liftetimes. In the present 
work, we extend the concept described by CW02 to inflow 
rates given by the model of Jupiter formation presented in 
Alibert et al. (2005b, hereafter referred A05) to calculate 
the structure of the Jovian circumplanetary disk which we assume 
to be well described by a two-dimensional time dependent 
turbulent $\alpha$-model.  Moreover, for the sake of 
consistency, we use in this calculations the same opacity law 
and equation of state as in A05. This allows us to reduce
quantitatively the number of free parameters
in our model of subnebula.

The evolution of the subnebula, which is strongly coupled to the last 
stages of the host planet formation, proceeds in two distinct phases. 
The first phase corresponds to the time during which the subnebula 
is still fed by the solar nebula which has not yet fully dissipated. 
Once this has occurred, the subnebula enters the second phase during 
which it evolves as a closed system until all the mass has been accreted 
by the planet.

Considerations about the thermodynamical state of the Jovian 
circumplanetary disk as well as about migration of proto-satellites 
in the subnebula allow us to discuss the survival of the Galilean 
satellites. We demonstrate that, during the beginning of phase 1 
(when the subnebula is fed by the nebula), proto-satellites cannot 
survive and fall onto Jupiter as a result of type I and type II migration.
Once the feeding of the subnebula ceases (phase 2), we show that 
satellites with properties quite similar to the Galilean ones can survive for
a limited range of the dissipation parameter $\alpha$.

The outline of the paper is as follows. In Sect. 2, we provide some 
details on the formation of Jupiter and its subnebula. We also describe 
the structure and the evolution of the circumplanetary disk.  In Sect. 3, 
we examine the conditions of migration of proto-satellites in the 
subnebula, and, in Sect. 4, we calculate the thermodynamical 
conditions inside the subnebula. Finally, Sect. 5 is devoted to the 
discussions and the summary. The composition of ices incorporated 
in the regular icy satellites of Jupiter, which derives from the 
thermodynamical conditions of the subnebula, will be discussed 
in a forthcoming paper (paper II). 

\section{Evolutionary turbulent model of the Jovian subnebula}
\subsection{Formation of Jupiter and its circumplanetary disk}

The model of the Jovian subnebula we consider here is derived
from the concept advocated both by Stevenson (2001) and CW02:
The so called gas-starved disk model.
We recall only briefly this concept as more detailled explanations can be
found in CW02.

At the begining of Jupiter's formation process, the total radius of the planet equals
its Hill's radius, and no subnebula can exist. 
At the end of the formation process, the cooling rate of the planet's envelope
is such that its radius shrinks faster than the disk can supply additional gas.
Accretion of gas now proceeds through streamers which eventually collide and
form a circumplanetary disk (Lubow et al. 1999).
When the radius of proto-Jupiter becomes small enough, a subnebula emerges from the
contracting atmosphere. This subnebula is fed by gas and gas-coupled solids accreted from the nebula.
Note however that, in the original model of CW02, the accretion rate from the solar
nebula to the subnebula remains a free parameter. In the model we present here,
we use the accretion rate of gas derived from the model of Jupiter formation
of A05. 

This formation model takes into account three effects, namely the formation process of the planet itself, 
by accretion of solids and gas, the protoplanetary disk structure and evolution,
and the migration of the planet. In this model, the giant planet forms from an embryo
originally located between 9 and 10 AU 
that ends its migration when the disk has disappeared, at the current position
of Jupiter. More informations about this model can be found elsewere (A05), and we only 
give here some details on the end of the formation process, when the subnebula
can exist.

During the formation of Jupiter, the planet accretes some
gas from the solar nebula.
The accretion rate of gas increases with time, and
is first governed by the internal
structure of the planet (which controls the amount of gas that can be accreted).
After $\sim 1.5$ Myr (see A05), the planet has opened a gap inside the disk, and the amount of gas that can be accepted
by the planet exceeds what can be sustained by the protoplanetary disk. The accretion
rate is then controlled by the solar nebula. Hydrodynamical simulations by Lubow et al. (1999)
and D'Angelo et al. (2002) have shown that the accretion rate of gas through a gap depends on
the mass of the planet and the characteristics of the disk. 
Analytical fits of the afore-mentionned simulations by Veras \& Armitage (2004)
lead to an accretion rate approximately given by:
\beq
{\dot{M}_{\rm gas,Max} \over \dot{M}_{\rm disk} }= 1.668 \left( \Mplanet / M_J \right)^{1/3}
e^{- {\Mplanet \over 1.5 M_J}} + 0.04,
\label{eq_max}
\eeq
where $M_J$ is the Jupiter mass and $\dot{M}_{\rm disk}$ is the accretion rate in the unperturbed solar nebula,
at the position of the planet. In our model, this gives, together with the value of $\dot{M}_{\rm disk}$ derived from the disk
model of A05, the amount of material flowing in the subnebula.

\subsection{Structure and evolution of the Jovian subnebula}

In our model, the evolution of the subnebula proceeds in two phases. During phase 1,
when the solar nebula is still present, the subnebula is fed through its outer edge, at a rate
ruled by Jupiter's formation (see below).
The outer radius of the subnebula is taken equal to 150 $R_J \sim 1/5 \times R_{\rm Hill}$,
a value close to the one calculated by Magni \& Coradini (2004) using a 3D
hydrodynamical model of the end of Jupiter's formation.

In phase 2, namely when the
solar nebula has disappeared, the subnebula evolves only due to accretion of its material onto the planet.
In addition, due to the conservation of angular momentum, the subnebula expands outward, its
outer radius being limited to the Hill's radius.
In both cases, the inner radius of the subnebula is chosen equal to 3 $R_J$,
a value lower than the Jovianocentric distance of Io, the innermost regular satellite of Jupiter.

\subsubsection{Temporal evolution}
\label{subdisk}

In the $\alpha$-disk model, the radial evolution of the subnebula is governed by a diffusion equation:
\beq
{d \Sigma \over d t} = {3 \over r} {\partial \over \partial r } \left[ r^{1/2} {\partial \over \partial r}
\nu \Sigma r^{1/2} \right]
\label{eq_diff_sub}
,\eeq
where $\Sigma$ is the surface density in gas phase in the subnebula, $r$ is the distance to the planet and $\nu$ the 
mean (vertically averaged) viscosity.
In this equation, the origin of time in the subnebula is arbitrarily taken when the accretion rate 
from the solar nebula to the subnebula is equal to  $9 \times 10^{-7} M_J/$yr. This corresponds
to the time when Jupiter has already accreted $\sim 75 \%$ of its total mass. However,
as we point out in Sect. \ref{thermo}, phase 1 of the subnebula's evolution is a
stationnary one, the accretion rate being constant for all radii. Hence, the origin of 
time has no great importance.

The mean viscosity used in the diffusion equation is determined with the help of the vertical
structure of the subnebula, calculated using the method described in Papaloizou \& Terquem (1999) 
and Alibert et al. (2005a). For each distance $r$ to the planet, the vertical structure is calculated by solving 
the equation for hydrostatic equilibrium, together with the energy equation and the diffusion 
equation for the radiative flux. The local viscosity is calculated using the standard Shakura \& 
Sunyaev (1973) $\alpha-$parametrization $\nu = \alpha C_s^2 / \Omega$ where the speed of sound 
$C_s^2$ is determined by the equation of state, and $\Omega ^2 = G \Mplanet / r^3$. Using this 
procedure, we calculate, for each distance $r$ to the planet and each value of the surface density, 
the distribution of pressure, temperature and density. We then derive the midplane pressure and 
temperature, as well as the mean viscosity inside the disk. This procedure avoids the use of an
{\it a priori} temperature (or viscosity) law as a function of the Jovianocentric distance.

From these calculations we are able to derive relations between the midplane pressure, 
temperature and surface density, that will be used to calculate the ices content of planetesimals 
accreted by the proto-satellites (see Sect. \ref{migration}).

\subsubsection{Boundary condition}

Eq. (\ref{eq_diff_sub}) is solved with two boundary conditions. At the inner boundary, we use the condition
\beq
\left. r \dpartial{ \nu \Sigma}{r} \)_{\rm inner \,\,\, radius} = 0
\label{BC_disk_int}
,\eeq
which states that the inner disk instantaneously adapts to the condition given by the outer disk (see Alibert et al. 2005a).

The outer boundary condition varies, depending upon the evolution of the subnebula (phase 1 or phase 2).
During phase 1, the subnebula is fed from its outer radius by gas originating from the solar
nebula. We then impose the gas accretion rate at the outer radius (see Alibert et al. 2005a):
\beq
\Phi(r) \equiv 3 \pi \nu \Sigma + 6 \pi r \dpartial{\nu \Sigma}{r} = {d M_{\rm J,gas} \over dt},
\eeq
where $d M_{\rm J,gas} / dt$ is the accretion rate of gas calculated in the Jupiter's formation model
and is given by Eq. (\ref{eq_max}).
In this equation, the accretion rate 
in the solar nebula away from the planet, is approximatly given 
by the following linear law\footnote{ 
The surface density in the protoplanetary disk,  $\Sigma_{\rm disk}$, evolves due to viscosity and photoevaporation. At the end of the protoplanetary
disk's life, the viscosity term ($\propto \nu_{\rm disk} \Sigma_{\rm disk}$) becomes negligible compared to the photoevaporation
term ($\propto \Sigma_{\rm disk} ^ 0$), leading to a quasi linear decrease of $\Sigma_{\rm disk}$ and of the accretion rate
($\dot{M}_{\rm disk} \propto \nu_{\rm disk} \Sigma_{\rm disk}$).
}:
\beq
\dot{M}_{\rm disk} \sim \dot{M}_{\rm disk,0} \left( 1 - t / \tau \right) \msol / {\rm yr}
\label{acc_rate}
,\eeq
where $\dot{M}_{\rm disk,0} \sim 1 \times 10^{-6} M_J /$yr and $\tau \sim 5.6 \times 10^5$ yr.

During phase 2, the subnebula is allowed to expand and the  surface
density is set to 0 at $\sim 700 R_J$ (corresponding to Jupiter's Hill radius).

\begin{figure}
\begin{center}
\epsfig{file=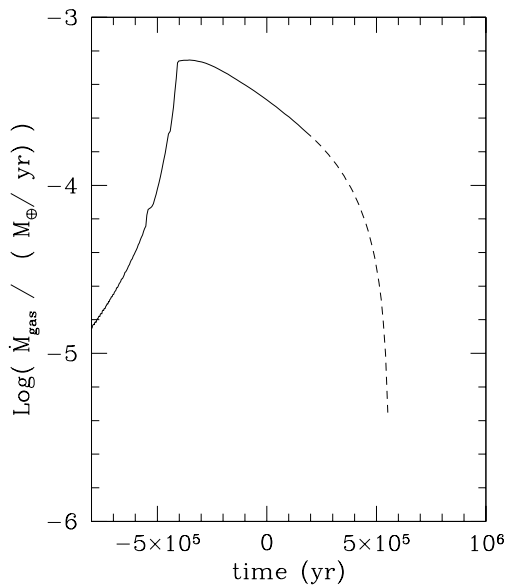,height=80mm,width=80mm}
\end{center} 
\caption{Gas accretion rate during the formation of Jupiter. The dashed line shows the fit we use in the calculation
of the subnebula. This analytical fit is calculated by combining Eq. (\ref{eq_max}) and Eq. (\ref{acc_rate}).
Jupiter's formation process starts $\sim 2$ Myr before the origin of time.}
\label{mdotgaz}
\end{figure}

\section{Proto-satellites migration}
\label{migration}

\subsection{Migration rates}
\label{mig_rates}

The forming satellites interact with the surrounding gas of the subnebula leading to migration.
In the present work, two mechanisms have to be considered, namely momentum exchange
via gravitational forces and gas drag.

Gravitationnal interactions with the subnebula lead to inward migration of the forming body. Depending on its mass,
two cases have to be distinguished. A low mass object migrate in what is called type I migration (Ward 1997),
whereas an object massive enough to open a gap in the subnebula migrates at a slower rate,
in type II migration (Ward 1997). The calculation of type I migration rate has been the subject of a lot of
studies, in particular in the context of planet formation. Analytically, the first migration rates
derived by Ward (1997) were so short that survival of planets was, contrary to what is observed,
very unlikely. Tanaka et al. (2002) have recently revisited these calculations, and found much longer
migration timescales, but still too short to ensure survival of forming protoplanets.
Numerically, Bate et al. (2003) considered a laminar disk (as was assumed in the work
of Tanaka et al. 2002) and found migration rates in good agreement with this latter work. On the other hand,
by considering protoplanetary disks with magnetic field, Nelson \& Papaloizou (2004) and Laughlin et al. (2004)
found that, due to density
fluctuations inside the disk, the migration pathway in type I can be represented by a random walk, reducing significantly
the net inward migration rate
compared to the one derived in Tanaka et al. (2002) and Bate et al. (2003). Moreover, type I migration rate
is very sensitive to local structures in the disk: Menou \& Goodman (2004) have shown that its rate
can be reduced by one order of magnitude in the regions of opacity transitions.
Finally, we note that in order to form planets, A05 had to drastically reduce the type I migration rate.

In the context of satellite formation in circumplanetary disks,
the kind of phenomenons that could reduce type I migration in
protoplanetary disks (MHD, density fluctuations, ...) may or may not  be present.
Moreover, we note that, 
in the model of A05, Jupiter and Saturn form
in a disk whose mass ratio (mass relative to the central body's one)
is of the order of $\sim 0.05$, whereas, as we shall see in Sect. \ref{thermo},
the mass ratio of the subnebula is of the order of $\sim 0.005$.
Since higher (relative) mass disks suffer some higher magnitude density
fluctuations, 
type I migration may be closer to the one of a laminar disk (Tanaka et al. 2003, Bate et al. 2003)
in the subnebula.

When the Hill's radius of the satellite (in the planet-satellite system) becomes higher than the local thickness of
the subnebula (calculated using the vertical structure, see Sect. \ref{subdisk}), 
migration switches to type II\footnote{ Note that the transition
from type I to type II could occur for much lower masses (see Rafikov 2002).}.
In that case, we use the same prescription as in Alibert et al. (2005a), namely :
\beq
{d \as \over d t} = - {3 \nu \over 2 \as} \times {\rm Min} \left( 1, {2 \Sigma \as ^2 \over \ms} \right)
,\eeq
where $\as$ is the jovianocentric distance of the satellite, and $\ms$ is the satellite mass.
In that case, the migration rate is given, for low mass objects, by the inward viscous velocity ($3 \nu / 2 \as$).
When the mass of the satellite becomes comparable to the one of the subnebula, migration slows down and eventually stoppes.

Two other phenomenons may lead to radial migration of forming satellites, namely gas drag and tidal
interactions with Jupiter. We have checked that
for the subnebulae we consider, and for bodies of mass close to the ones of Jovian satellites,
migration occurs mainly in type I, and, to a lower extent, in type II. Gas drag and tidal forces are negligible,
at least for the jovianocentric radii we consider.

\subsection{How to choose the dissipation parameter $\alpha$ ?}

The thermodynamical conditions inside the subnebula determine the composition
of ices incorporated into icy planetesimals (see paper II), which ultimately
determine the composition of the Jovian satellites. Hence, the modelling of the
thermodynamical evolution of the subnebula must be carried out in a way
compatible with the bulk chemical composition of the Jovian satellites.
Since in our model this thermodynamical evolution is mainly governed by the
$\alpha$-parameter, the bulk composition of the satellites provides constraints on the
allowed values for $\alpha$. We note that since the evolution of the subnebula is 
coupled to Jupiter's formation model, the constraints obtained for $\alpha$ may be
functions of this model as well. While the decrease of the mass inflow rate is important,
tests have shown that the timescale over which this happens does not seem to be crucial
as long as reasonable values are chosen.

During phase 1 the subnebula is fed through its outer edge by the nebula
at a rate which is determined by the parameters adopted for Jupiter's formation model.
The choice of $\alpha$ in the subnebula will determine how efficiently this inflow
will be processed through the subnebula. Clearly, lower values of $\alpha$ imply lower
transport rates and consequently higher surface densities for a fixed mass inflow. Since the
rate of type I migration is a function of surface density and not viscosity,
lower values of $\alpha$ also imply a larger migration rate. Hence, subnebulae
characterized by low $\alpha$ values are more massive, and imply  a faster type I
migration rate.
Moreover, a low viscosity subnebula being more massive, it is thicker.
Since the transition from type I to type II occurs  when
the Hill's radius of the satellite becomes greater than the subnebula's half-thickness,
the type I migration extent is higher in a low viscosity subnebula.

For type II migration, the rate is proportional to $\alpha$ as
long as the satellite mass remains small compared to the disk mass.
However, since the disk mass is larger for low $\alpha$ values, type II migration
in low $\alpha$ disks lasts longer than in high $\alpha$ disks.
This effects offsets, at least in part, the reduced migration rate in a low
$\alpha$ disk.

As we shall see in the next section, due to the afore-mentionned effects,
the distance a satellite migrates (due to both type I and type II migrations)
is larger in a low $\alpha$ disk.

With time, the temperature and pressure conditions at a given radius inside the subnebula decrease. Rocky satellites
must then form during the early times of the subnebula's evolution,
and, as a consequence, they suffer a large amount of migration. 
A low viscosity subnebula then looses all its rocky satellites inside the central body.
In order to allow the survival of these satellites, the dissipation
parameter $\alpha$ has to be greater than
a minimum value that is determined in the next section.

To obtain a maximum value of the dissipation parameter $\alpha$, we 
consider the formation of Callisto. Due to its partial differentiation, it is generally
accepted that this
satellite has been formed on a relatively long timescale (see CW02,
Mosqueira \& Estrada 2003a for discussions), of the order of at least $\sim 10^{5}$ years.
Since Callisto incorporates a large amount of ices (Sohl et al. 2002), its formation requires that
the subnebula remained cold in regions beyond $26.6 R_{\rm J}$ (its present day Jovianocentric
distance) during more than $\sim 10^{5}$ years, if one assumes that the formation
of the satellites has to be completed before the subnebula has disappeared.

During nearly all phase 1 of the subnebula's evolution, the temperature remains high due to the
high surface density. The temperature generally allows the presence of ices only after $\sim 0.5$
Myr (corresponding to the switch from phase 1 to phase 2, see Sect. \ref{thermo}). 
Whereas the dissipation parameter does not strongly influence the potential presence of ices during phase 1,
its value determines the evolution timescale of the subnebula during phase 2.
The higher $\alpha$, the shorter evolution timescale. Requiring that the temperature
must allow the presence of ices during at least $10^5$ yr translates in a maximum
value of the dissipation parameter $\alpha$.

\subsection{Migration pathways}

We assume that the satellite's mass increases linearly, on a timescale $\tau_S$, until it reaches a mass $M_S$.
The Jovianocentric distance decreases as a result of type I and/or type II migration.
The type I migration rate is calculated by using the formula of Tanaka et al. (2002), reduced by
a constant factor $f_I$ that can take two values: $f_I$ = 1 (no reduction) or $f_I$ = 0.001 (the value
used in the Jupiter's formation model of A05). As stated in Sect, \ref{mig_rates}, these two choices reflect
our lack of knowledge of the actual type I migration rate in the subnebula.

The proto-satellites start their drift at various locations and epochs, and their migration is calculated either
until they reach the inner edge of the subdisk, or until the subnebula has disappeared.
Proto-satellites whose jovianocentric distance becomes lower than the
inner radius are considered to have been lost inside the planet.

At each timestep, we calculate the local disk temperature and assume two planetesimals compositions
 that depend only on this thermodynamical parameter. For temperatures lower than 150 K, corresponding
to the condensation temperature of water, we assume icy planetesimals (with a given ices/rocks (I/R) ratio).
 In the opposite case, planetesimals are made of pure rocks.
At the time the subdisk disappears, we calculate the location of the remaining satellites,
and  the fraction of icy planetesimals accreted. For a given I/R ratio,
we can finally calculate the ices content of the satellites. Note however that the results presented  
in Figs. \ref{Io_2} to \ref{Cal_4} do not depend on
the assumed I/R ratio of planetesimals.

We have made simulations with $M_S$ equal to the masses of the two rocky Jovian satellites,   Io and Europa
 (Europa, with an ices content lower than 10 \% (Sohl et al. 2002), is rather considered as a rocky satellite),
and the mass of Callisto.
The formation timescale $\tau_S$ can take different values, between $10^4$ and $10^6$ yr for the rocky
satellites, and between $10^5$ and $10^6$ yr for Callisto.
 Note that we do not adress in this paper the question of the possibility of forming a body of mass $M_S$ on a
timescale $\tau_S$. 
We assume in these simple calculations that the formation efficiency does not vary with time. Our calculations then allow
 us to determine the distance to Jupiter where satellites {\it can} exist when the subnebula as cleared,
 as a function of their ices content.

\subsubsection{Rocky satellites}

For $f_I = 0.001$ (type I migration rate equal to the one in A05), rocky satellites can survive 
for whatever value of $\alpha$ we have considered (down to $10^{-6}$). No constraining minimum value of the
dissipation parameter can be derived in this case.

\begin{figure}
\begin{center}
\epsfig{file=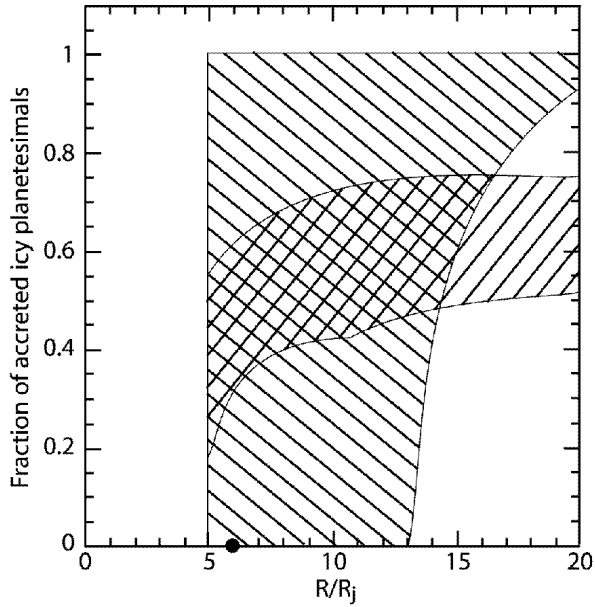,height=80mm,width=80mm}
\end{center}
\caption{Composition (Fraction of accreted icy planetesimals)
 {\it vs} jovianocentric distance for a set of simulations. The
mass of the satellite is the one of Io ($0.47 \times 10^{-4} M_J$) and the forming time $\tau_S$ is equal to $10^5$ yr ($\backslash
\backslash \backslash$)
and $10^6$ yr ($///$). The result for $\tau_S = 10^4$ yr is very close to the one for $\tau_S = 10^5$ yr and is not
 represented. The $\alpha$-parameter is equal to $2 \times 10^{-4}$. The big black dot represents the position of Io in
 this diagram, with an I/R ratio equal to 0 (Sohl et al. 2002). The type I migration rate is calculated using the formula derived
by Tanaka et al. (2002) without modification.}
\label{Io_2}
\end{figure}

We now focus on $f_I=1$ (no reduction of type I migration).
Figure \ref{Io_2} gives, as a function of the distance to the planet, the relative mass of ices inside
the satellite, for $\alpha = 2 \times 10^{-4}$. The assumed formation
time is varied from $10^4$ to $10^6$ yr, and the mass of the satellite is the one of Io.
Rocky satellites can be found at locations up to $\sim 13 R_J$. On the opposite,
for values of $\alpha$ lower than this minimum value ($10^{-4}$ and $1.5 \times 10^{-4}$ have been
tested), 
no rocky satellite of the mass of Io can be formed and survive at its present
location ($\sim 6 R_J$).
If the mass of the satellite is the one of Europa, the minimum value of $\alpha$ is similar.

This result does not depend on the formation timescale provided it is below $10^5$ years.
On the opposite, if $\tau_S = 10^6$ yr, purely rocky satellites cannot be found at the actual location of Io or Europa.
This is expectable since, in our model, the condensation radius has reached the inner location of the subnebula in $\sim 0.7$ Myr.
However, this formation timescale seems too large for the accretion of Galilean satellites.

We conclude that, if type I migration is not reduced,
the $\alpha$ parameter has to be chosen greater than a 
minimum value, of the  order of 2 $\times 10^{-4}$.

Finally, note that the bodies whose final properties ressemble those of Io have started their formation
during the end of phase 1. On the other hand, as pointed out by CW02, satellites formed during the
early times of the subnebula's evolution (beginning of phase 1, corresponding to high inflow rates) experience very
strong disk torques, leading to very rapid inward migration:
they end their migration below the inner radius of the subnebula,
regardless the value of $\alpha$ considered.
If the tidal forces between the planet and the satellite are not strong enough to prevent further migration,
these satellites will be
accreted by the planet.
These satellites formed at the beginning of phase 1, may therefore have
contributed to the enrichment of the giant planet in heavy elements.

\subsubsection{Icy satellites}
\label{icy}

We now consider the formation of Callisto, and use $f_I=1$ to calculate the type I migration rate (however,
the results do not strongly depend on the actual value of $f_I$).

\begin{figure}
\begin{center}
\epsfig{file=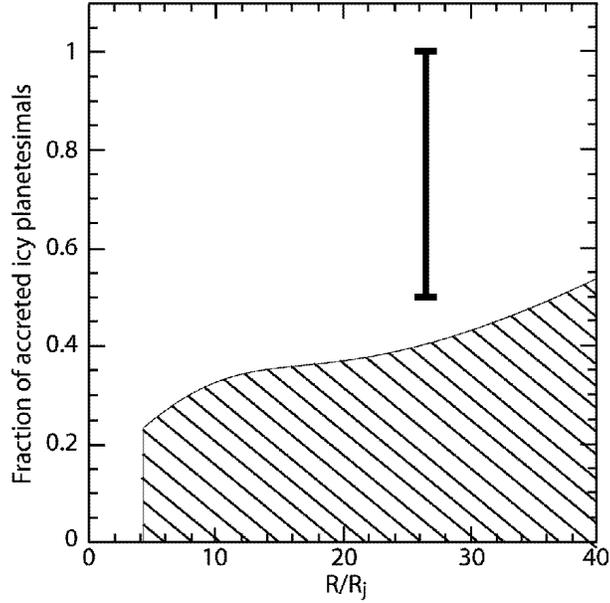,height=80mm,width=80mm}
\end{center}
\caption{Same as Fig. \ref{Io_2}, but for 
the mass of Callisto ($0.57 \times 10^{-4} M_J$).
 The $\alpha$-parameter is equal to $2 \times 10^{-3}$. The "error bar" represents the position of Callisto in
 this diagram, with an ices content of $\sim 40 {\rm wt} \%$ to $\sim 60 {\rm wt} \%$ (Sohl et al. 2002), and assuming a mean
I/R of 1 to 4 in the accreted icy planetesimals.}
\label{Cal_3}
\end{figure}

\begin{figure}
\begin{center}
\epsfig{file=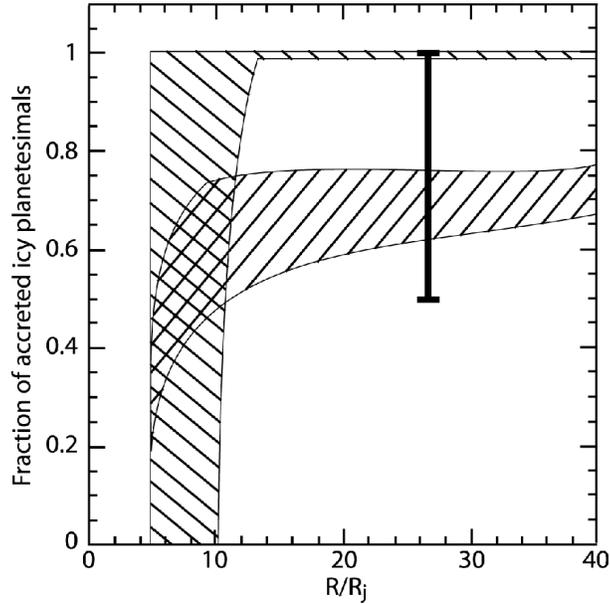,height=80mm,width=80mm}
\end{center}
\caption{Same as Fig. \ref{Cal_3}, but for $\alpha = 2 \times 10^{-4}$.
The forming time $\tau_S$ is equal to $10^5$ yr ($\backslash
\backslash \backslash$) and $10^6$ yr ($///$).}
\label{Cal_4}
\end{figure}

Figures \ref{Cal_3} and \ref{Cal_4} give, as a function of the distance to the planet, the relative mass of ices inside
the satellite, for the dissipation parameters $\alpha = 2 \times 10^{-3}$ and $\alpha = 2 \times 10^{-4}$. 
The formation timescale $\tau_{\rm S}$ is equal to $10^5$ and $10^6$ yr for $\alpha = 2 \times 10^{-4}$,
and $10^5$ y for $\alpha = 2 \times 10^{-3}$.
In this latter case, at the present location of Callisto, satellites incorporate less than $\sim 40 \%$ 
of icy planetesimals. In order to explain the actual I/R in Callisto (at least 40 wt\% of ices, see Sohl et al. 2002),
one has to suppose that icy planetesimals contain no rocks, which is unlikely. If the formation timescale is increased
to 1 Myr, no icy satellites can form and survive at the present location of Callisto. This results from the very rapid
evolution of the subnebula in phase 2: ices do not exist long enough to allow the formation of icy satellites
within $10^6$ years.

On the other hand for $\alpha = 2 \times 10^{-4}$ and using the two afore-mentionned formation timescales, it is possible to
find satellites that have accreted at least 60 \% of icy planetesimals. This is compatible with
the ices content estimation of Sohl et al. (2002), if the I/R ratio of planetesimals is a least equal
to $\sim 2$.
The results are similar for $f_I=0.001$. This is to be expected since
the important parameter is the
evolution timescale of the subnebula, governed by the value of $\alpha$.
Hence, whatever the value of $f_I$, the dissipation parameter
has to be lower than $2 \times 10^{-3}$ in order to account for the ices content and
formation timescale of Callisto. 

In summary, the ices content and formation time of Callisto require a slow dissipating nebula,
which translates in a dissipation parameter
lower than $\sim 10^{-3}$. Moreover, the survival of rocky satellites requires a subnebula not too massive.
This 
gives a lower value of the dissipation 
parameter, that depends on the type I migration rate. For a low migration rate ($f_I=0.001$), 
there is no constraining minimum value, whereas, for $f_I=1$ (migration rate as derived by Tanaka et al. 2002), the minimum
value is of the order of $2 \times 10^{-4}$.

These boundaries on the $\alpha$ parameter may depend on the Jupiter formation model,
which has an effect on the inflow rate, through the evolution of the planet's
mass (see Eq. (\ref{eq_max})) and the evolution of the accretion rate
inside the protoplanetary disk ($\dot{M}_{\rm disk}$, given by Eq. (\ref{acc_rate})).
In order to quantify this effect, we have varied the parameter $\tau$ in Eq. (\ref{acc_rate}),
from 0.1 Myr to 1 Myr, keeping the total mass processed inside the subnebula constant.

The minimum value of $\alpha$ is found between $10^{-4}$ and $2 \times 10^{-4}$ for 
all the cases considered, whereas the maximum value can be increased to  $\sim 3 \times 10^{-3}$
for $\tau = 1$ Myr\footnote{ 
We note moreover that, as discussed in CW02, this latter value of the subnebula lifetime is hardly compatible
with the present obliquity of Jupiter.
}. We then conclude that the boundaries we obtain do not strongly
depend on the Jupiter formation model. 
Finally, the variation of the boundary values of $\alpha$ are found to be
similar using the criterion given in Rafikov (2002)
to convert from type I to type II migration rate.

\section{Thermodynamical conditions inside the subnebula}
\label{thermo}

Once the outer radius of the subnebula and the accretion rate from the nebula have been
fixed (see sections above), the structure of the Jovian subnebula only depends on $\alpha$.
The structure of the subdisk has been calculated for a large range of $\alpha$, namely 
between $10^{-5}$ and $10^{-2}$, and we present here the resulting model for 
$\alpha = 2 \times 10^{-4}$. As shown in the previous section, this value leads to
a subnebula's structure compatible with the presence of rocky satellites at their current 
locations (whatever the assumed value of the type I migration rate we have considered).
This subnebula's structure is also compatible with the low formation process of Callisto,
and its ices content.

\begin{figure}
\begin{center}
\epsfig{file=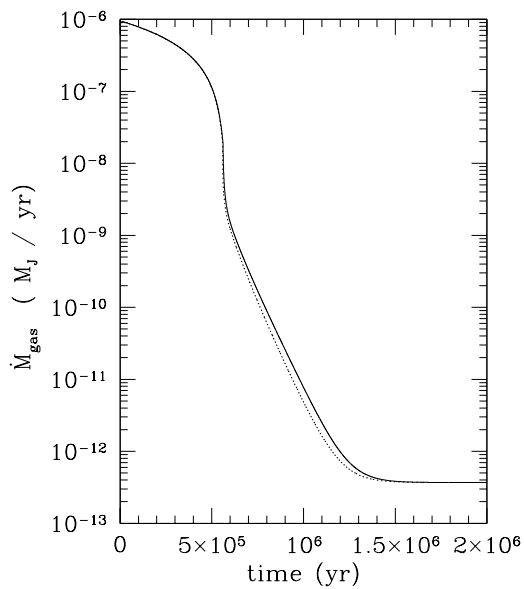,height=80mm,width=80mm}
\end{center}
\caption{Gas accretion rate at the inner edge of the subdisk (solid line), and at $150 R_J$ (dotted line)
(outer edge of the subdik during phase 1) as a function of time in the Jovian 
subnebula. The transition from phase 1 to phase 2 occurs at $t$ = 0.56 Myr. 
Before this time, the accretion rate is nearly constant (as a function of the jovianocentric distance) in the whole 
subnebula, whereas it varies during the subsequent evolution.}
\label{macc}
\end{figure}

Figure \ref{macc} gives the evolution of the accretion rates at both the inner edge of the subnebula,
and  at $150 R_J$ (the position of the outer edge during phase 1). 
Phase 1  lasts until the protoplanetary disk has disappeared, namely at $t$ = 0.56 Myr 
according to our model\footnote{However, as stated in Sect.
\ref{subdisk} the origin of time is quite arbitrary}. During phase 1, the two accretion rates are nearly identical,
revealing that the subnebula remains in equilibrium. After $t$ = 0.56 Myr, the 
subnebula enters in its second phase of evolution
and begins its expansion. The accretion rate as a function of the distance to Jupiter 
is then no longer constant. Hence, one cannot use an equilibrium model ({\it i.e.} 
with a constant accretion rate throughout all the subnebula)
for this phase, and the diffusion equation (Eq. (\ref{eq_diff_sub})) has to be solved.

\begin{figure}
\begin{center}
\epsfig{file=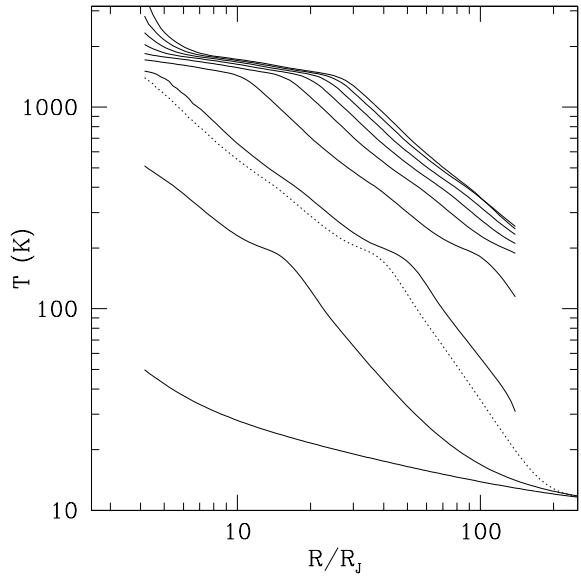,height=80mm,width=80mm}
\end{center}
\caption{Log of temperature at different epochs in the midplane of the Jovian subnebula, at times
(from top to bottom) 
$t$ = 0, 0.1 Myr, 0.2 Myr, 0.3 Myr, 0.4 Myr, 0.5 Myr, 0.56 Myr, 0.6 Myr and 1 Myr.
The dotted line gives the temperature $2 \times 10^3$ yr after 0.56 Myr (the epoch from which the
outward diffusion of the Jovian subnebula begins). For $t$ $\leq 0.56$ Myr, the outer radius 
of the subnebula is equal to 150 $R_J$}
\label{Tdisk}
\end{figure}

\begin{figure}
\begin{center}
\epsfig{file=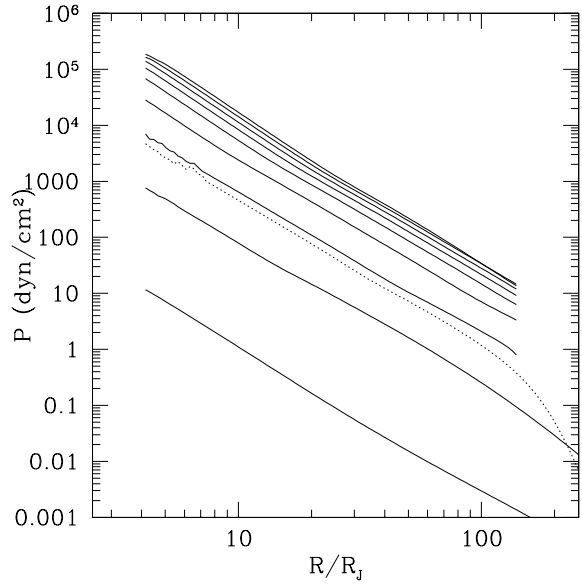,height=80mm,width=80mm}
\end{center}
\caption{Log of pressure at different epochs in the Jovian subnebula midplane. Times are the same as in Fig. \ref{Tdisk}.}
\label{Pdisk}
\end{figure}

\begin{figure}
\begin{center}
\epsfig{file=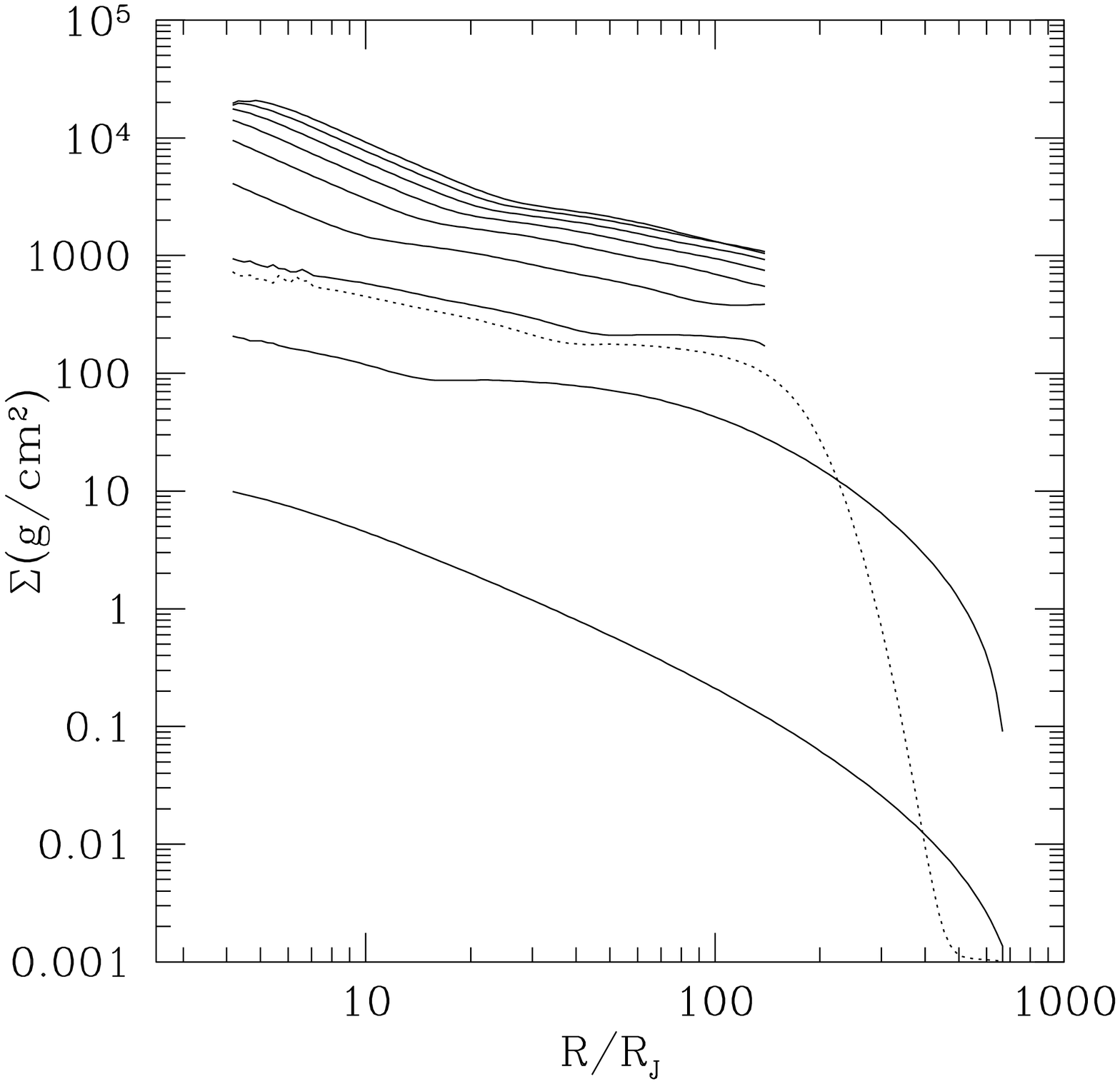,height=80mm,width=80mm}
\end{center}
\caption{Log of surface density at different epochs in the Jovian subnebula midplane.
Times are the same as in Fig. \ref{Tdisk}.}
\label{Sigdisk}
\end{figure}

Figures \ref{Tdisk} to \ref{Sigdisk} show radial profiles of temperature $T$, pressure $P$, and surface density $\Sigma$, 
 respectively, at various epochs. Water is in vapor phase at $t$ = 0 in the whole subnebula, and $T$, $P$, and $\Sigma$
 decrease with time and with the distance to Jupiter. At t = 0.44 Myr water starts to crystallize at the outer edge of 
the disk. The cooling of the Jovian subnebula results in the inward propagation of a water condensation front 
reaching the orbits of Callisto (26.6 $R_{J}$), Ganymede (15.1 $R_{J}$), Europa (9.5 $R_{J}$) and Io (6 $R_{J}$)
 at $t$ = 0.57 Myr, 0.61 Myr, 0.69 Myr and 0.77 Myr, respectively. Note that only 0.2 Myr are required for
 the snow line to propagate between the orbits of Callisto and Io. This very short timescale is due to the fact
 that crystallization of water occurs in the satellite zone during the transition between phases 1 and 2 of the subnebula
 evolution, when the cooling of the subnebula is very efficient.

       \begin{table}[h]
      \caption[]{Thermodynamical parameters of the Jovian subnebula (nominal model).}
      \begin{center}
         \begin{tabular}[]{lc}
            \hline
            \hline
            \noalign{\smallskip}
               Thermodynamical  &  \\
               parameters  & \\
             \noalign{\smallskip}
             \hline
             \noalign{\smallskip}
             Mean mol. weight (g/mole) &  2.4 \\
             $\alpha$  & $2 \times 10^{-4}$ \\
             Initial disk's radius ($R_{J}$) & 150 \\
             Initial disk's mass ($M_{J}$) & $3 \times 10^{-3}$ \\
             Initial accretion rate ($M_{J}$/yr) & $9 \times 10^{-7}$ \\
           \noalign{\smallskip}
            \hline
         \end{tabular}
   \end{center}
   \end{table}

\begin{figure}
\begin{center}
\epsfig{file=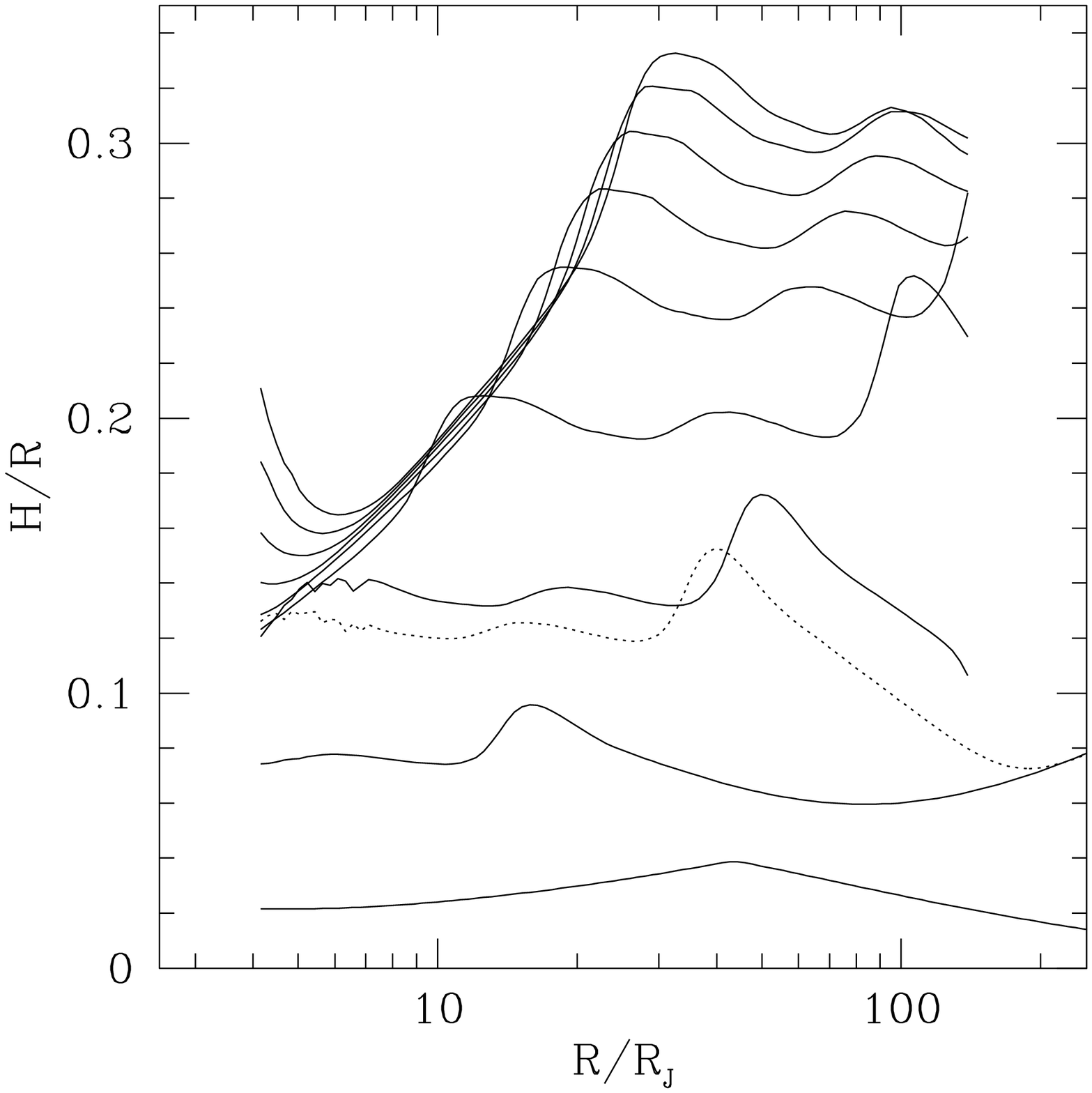,height=80mm,width=90mm}
\end{center}
\caption{Jovian subnebula $H/R$ profiles at different epochs. Times are the same as in Fig. \ref{Tdisk}.}
\label{Hdisk}
\end{figure}

Figure \ref{Hdisk} represents radial profile of the $H/R$ ratio throughout the Jovian subnebula at different epochs, 
where $H$ is the half-height of the disk and $R$ is the Jovianocentric distance. The value of $H/R$ is initially quite high,
 but it rapidly evolves to values low enough ($H/R$ $<$ 0.3 at $t$ = 0.2 Myr) to make the thin disk approximation a reasonable one.
 Moreover, $H/R$ rapidly decreases with time and becomes lower than 0.1 during the second phase of the subnebula evolution, when the solar nebula has vanished.

\section{Discussion and Summary}

We have presented in this paper a model of the Jovian subnebula
derived from the concept initially proposed by Stevenson (2001) and CW02, which is 
consistent with the formation of Jupiter, as described in A05. 
The evolution of the subnebula can be divided in two phases. 
During the first phase, the subnebula is fed through its outer 
edge by gas and gas-coupled solids originating from the protoplanetary 
disk.  During this phase, the subnebula is in equilibrium, its accretion 
rate beeing constant through the subdisk and driven by the inflow of 
material delivered by the protoplanetary disk. When the solar nebula 
disappears, the subnebula enters in its second phase of evolution. 
The mass flux through the outer edge stops, and the Jovian subnebula 
expands outward due to viscosity. The subnebula is then no longer in global 
equilibrium since the accretion rate varies with the jovianocentric 
distance in the subdisk. 

In order to determine the thermodynamical properties of the subnebula,
we had to
choose the viscosity parameter $\alpha$, in a way compatible with
the present day structure of the Jupiter's satellite system.

Using the type I migration rate derived by
Tanaka et al. (2002), migration calculations showed that too small viscosity parameters
are incompatible with the presence of rocky satellites at
distances up to $\sim 10 R_{\rm J}$. The value of
the minimum $\alpha$-parameter was found to be $\sim 2 \times
10^{-4}$.
A reduced migration rate, on the other hand, 
resulted in a very small minimum 
value of the $\alpha$-parameter. 

The formation timescale and ices content of Callisto gives an upper
limit on the dissipation parameter.
High values of $\alpha$ lead to a subnebula where ices can be present
only during a short time. This is due to the very rapid decrease of the
surface density during phase 2 of the subnebula's evolution.
In order to have ices present during at least $\sim 0.1$ Myr
(the minimum formation timescale of Callisto, see CW02,
Mosqueira \& Estrada 2003), the subnebula must evolve slowly
enough during phase 2. In the framework of our model, the $\alpha$-parameter
must be lower than $\sim 10^{-3}$. 
Tests have shown that the boundary values of alpha do not strongly
depend on the assumed inflow rate of material, and on
the Jupiter formation model.

Using a viscosity parameter compatible with these two boundaries,
we have finally derived the thermodynamical conditions inside the
subnebula, as a function of jovianocentric radius and time. 
We will study the implications of these results 
for the gas phase chemistries of carbon and nitrogen
and the derived composition of ices incorporated in the regular icy satellites
in a forthcoming paper (see paper II).

The calculations presented here are subject to some limitations.
First, our subnebula model is based on the Jupiter formation
model of A05, and one must keep in mind that the results depend on this
assumption. Second, as in A05, we use the $\alpha$ formalism
(Shakura \& Sunyaev 1973) to derive the subnebula's structure,
which itself is a limitation. Third, the type I migration rate is still
subject to a large uncertainty, which we parametrize in a crude way
by the factor $f_I$. Some more precise constraints on both the
viscosity and the migration processes could be
brought using more precise calculations of the formation of
satellites. This needs in particular to calculate the planetesimals surface density,
and must take into account the effect of gas drag and growth of solids by
mutual collisions.

Finally, our migration calculations also show that any satellite starting 
its formation during the beginning of the subnebula's
life is accreted onto Jupiter. 
The accretion of early formed satellites may be an alternative 
way to direct accretion by the planet, to increase the final
content of heavy elements of Jupiter. 

\begin{acknowledgements}
We thank C. Winisdoerffer for useful comments.
This work was supported in part by the Swiss National Science Foundation. OM was supported by an ESA external fellowship, and this support is gratefully acknowledged.
\end{acknowledgements}

\end{document}